\documentclass[journal]{IEEEtran}

\usepackage{cite}
\usepackage{hyperref}       
\usepackage{url}
\usepackage{booktabs}       
\usepackage{amsfonts}    
\usepackage{amssymb}
\usepackage{amsmath,bm}
\usepackage{graphicx}
\usepackage{caption}
\usepackage{algpseudocode, algorithmicx, algorithm,mismath}
\usepackage{siunitx}
\usepackage{multirow}
\usepackage{pdfpages}
\usepackage{svg}
\usepackage{changes}
\usepackage{colortbl}
\usepackage{breqn}
\newcommand \argmax {\operatorname*{argmax}}

\newcommand \andothers {\textit{et~al.}}

\newcommand \Fig {Fig.}
\newcommand \commentcolor {black}
\newcommand{\mathbold}[1] {\textbf{\textit{#1}}}
\ifCLASSINFOpdf
\else
\fi
\ifCLASSOPTIONcompsoc
\usepackage[caption=false,font=normalsize,labelfont=sf,textfont=sf]{subfig}
\else
\usepackage[caption=false,font=footnotesize]{subfig}
\fi

\ifCLASSOPTIONcaptionsoff
\usepackage[nomarkers]{endfloat}
\let\MYoriglatexcaption\caption
\renewcommand{\caption}[2][\relax]{\MYoriglatexcaption[#2]{#2}}
\fi

\usepackage{url}

\begin{document}
	
	\title{Online Electric Vehicle Charging Detection Based on Memory\-based Transformer using Smart Meter Data}
	
	\author{\href{https://orcid.org/0000-0002-7441-9344}{\includegraphics[scale=0.06]{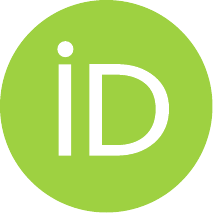}\hspace{1mm} Ammar Kamoona\thanks{This work is supported by Australian Research Council through
				project NO. LP180101309 and Victorian Government through Victorian
				Higher Education State Investment Fund (Supporting electrification of
				Victoria’s future fleet).\\
				A. Kammona, H. Song, M. Jalili and X. Yu are with the School of Engineering, Royal Melbourne Institute of Technology, Melbourne, Australia, e-mail: \{ammar.kamoona, hui.Song,,mahdi.jalili,xinghuo.yu, @rmit.edu.au\}. H. Wang and Reza Razzaghi are with Monash University, Australia.}},\hspace{1mm} Hui Song, ~\href{https://orcid.org/0000-0002-0517-9420}{\includegraphics[scale=0.06]{orcid.pdf}\hspace{1mm} Mahdi Jalili},\hspace{1mm} Hao Wang,\hspace{1mm} Reza Razzaghi, Xinghuo Yu }
	

%
%

\markboth{IEEE TRANSACTIONS class,~Vol.~14, No.~8, 2022}%
{Shell \MakeLowercase{\textit{et al.}}: Bare Demo of IEEEtran.cls for IEEE Journals}
%



\maketitle
\begin{abstract}
The growing popularity of Electric Vehicles (EVs) poses unique challenges for grid operators and infrastructure, which requires effectively managing these vehicles' integration into the grid. Identification of EVs charging is essential to electricity Distribution Network Operators (DNOs) for better planning and managing the distribution grid. One critical aspect is the ability to accurately identify the presence of EV charging in the grid. EV charging identification using smart meter readings obtained from behind-the-meter devices is a challenging task that enables effective managing the integration of EVs into the existing power grid. Different from the existing supervised models that require addressing the imbalance problem caused by EVs and non-EVs data, we propose a novel unsupervised memory-based transformer (M-TR) that can run in real-time (online) to detect EVs charging from a streaming smart meter. It dynamically leverages coarse-scale historical information using an M-TR encoder from an extended global temporal window, in conjunction with an M-TR decoder that concentrates on a limited time frame, local window, aiming to capture the fine-scale characteristics of the smart meter data. The M-TR is based on an anomaly detection technique that does not require any prior knowledge about EVs charging profiles, nor it does only require real power consumption data of non-EV users. In addition, the proposed model leverages the power of transfer learning. The  M-TR is compared with different state-of-the-art methods and performs better than other unsupervised learning models. The model can run with an excellent execution time of 1.2 sec. for 1-minute smart recordings.
\end{abstract}

\begin{IEEEkeywords}
Electric vehicle identification, online anomaly detection, transformer, smart meter data.
\end{IEEEkeywords}

%
\IEEEpeerreviewmaketitle

\section{Introduction}
%
%
%
%
\IEEEPARstart{}{}  
Transport electrification is critical to the transition to a more sustainable energy system. The electrification of transportation provides a powerful solution for reducing greenhouse gas emissions and mitigating the negative impacts of fossil fuel consumption as the world continues to struggle with climate change. About one-quarter of the energy-related emissions are estimated to be attributed to the transportation sector alone~\cite{yong2015review}. In recent years, the market has seen an increase in the adoption of electric vehicles (EVs) to decarbonize the transport sector. The emissions of CO2 produced by EVs are significantly lower than those produced by gasoline or diesel engines. Globally, approximately 100 million EVs will be on the road by 2035, according to the Energy Outlook~\cite{noauthor_prospects_nodate}. Different approaches are being used worldwide to encourage the uptake of EVs, including financial incentives, regulations, investment in charging infrastructure, and government fleet procurement, to reduce greenhouse emissions and support a sustainable and resilient transportation system.

Several works~\cite{zhang2018optimal,yong2015review,cao2019toward} have outlined that new technical challenges will arise with an increase in EV uptake in the power distribution grids in terms of power demand increase, system losses, voltage drops, phase unbalance, and stability issues. Addressing these challenges has led to a new research area to emerge with different approaches, mainly focusing on demand-response and public charging stations. A variety of approaches are being used, including coordinated EV charging using optimal charging price~\cite{zhang2018optimal}, and EV charging location planning based on competition in resource allocation data~\cite{YanEV2021}. Most of these studies assume that EV charging occurs mostly at public stations. Most EV owners, however, charge their EVs at home since they enjoy the convenience and flexibility of choosing when to charge and the lower electricity costs during off-peak periods. Furthermore, according to the current Australian study~\cite{mortimore2022business}, even 47\% of the business fleets are usually home-garaged. 

To manage the impact of EV charging at homes, distribution network operators (DNOs) and retailers are interested in mitigating the impact using smart charging and scheduling algorithms~\cite{lee2021adaptive}. However, online EV identification, i.e., detecting EV charging in a household using its streamed smart meter data, is the main prerequisite. Online EV charging identification can help DNOs  understand the impact of EVs on the local grid and identify opportunities for grid optimization. Streamed behind-the-meter data can help DNOs understand EV owners' charging behavior through EV identification and develop targeted interventions to encourage off-peak charging or manage peak demand.


EV identification can be categorized based on the availability of the training data into supervised, semi-supervised, and unsupervised approaches. The supervised learning uses EV and non-EV smart meter readings for training the machine learning model. However, due to the sparsity of EV charging sessions/events over time, this may suffer from data imbalance problems~\cite{HuiEV2021iit}. Different approaches can be used to overcome data imbalance, such as over-sampling~\cite{ye2020oversampling} or under-sampling~\cite{HuiEV2021iit}. But these methods could bring in new limitations. For example, under-sampling excludes the majority of samples, which can potentially reduce model performance. While, over-sampling requires a high computation cost, and the redundancy in the minor class may reduce model classification accuracy~\cite{HuiEV2021iit}. It is worth mentioning that in the semi-supervised learning, Jahangir~\andothers~\cite{jahangir2021novel} proposed to address the problem of different EV demand characteristics using 3-D convolution via a semi-supervised approach using Generative Adversarial Networks (GANs). 

The unsupervised learning techniques can provide a more suitable and efficient approach that does not require minor class data during the training. Different unsupervised have been proposed in the literature~\cite{wang2020deep,munshi2018unsupervised,xiang2021charging} for EV charging profile identification. Wang~\andothers~\cite{wang2020deep} proposed a deep generative model for EV charging profile extraction using Hidden Markov processes. The aforementioned studies required complete knowledge of EVs' arrival and departure times, but DNOs lack access to these data in most cases. Munshi~\andothers~\cite{munshi2018unsupervised} used Independent Component Analysis (ICA) to decompose the smart meter data and extract EV charging load pattern. However, their approach requires different extraction processes to identify different charging patterns, which is very time-consuming. Xiang~\andothers~\cite{xiang2021charging} proposed training free charging load profile extraction by applying two-stage signal decomposition. However, the earlier approach is based on the assumption that the EV load profile is very distinctive from others. They considered charging characteristics, such as the EV power consumption and the rectangle profile of charging/discharging, are known. In practice, information received by the energy distributors regarding a new EV charging occurrence can be incomplete, and lagging and charging profiles can not be generalized to all cases with different EV models and in different EV charging stations. More importantly, DNOs cannot access information about charging profiles, arrival time, and departure time of EVs. Another approach that does not require prior knowledge about the charging profile is anomaly detection, a branch of unsupervised learning.

Anomaly detection, which is the main focus of this paper, can be defined as detecting instances in the data that deviate from the predefined normal model~\cite{chandola2009anomaly,8986829}.
The application of anomaly detection ranges from  security, risk management, health, and medical risks~\cite{chandola2009anomaly}. 
Most of EV identification studies aim at finding an EV charging session that is highly based on the assumption that an EV presence in the household. However, in our problem setting, we do not rely on such assumption.  The aim of our paper is to introduce an online machine-learning model capable of detecting EV charging from behind the meter in an unsupervised fashion.

Inspired by the success of Transformers~\cite{vaswani2017attention} in different domains, particularly in natural language processing, this paper utilizes Transformer for unsupervised online EV charging identification. Transformer  has shown superior performance in modeling long-term dependencies of sequential data compared to Recurrent Neural Networks (RNNs). Different from other research, in this paper, we propose an online memory-based transformer (M-TR) for unsupervised EVs charging identification using streamed smart meter data. The contributions of the paper are summarized as follows: 
\begin{itemize}
\item We propose an online unsupervised learning framework for online EV charging identification using smart meter data that does not rely on the assumption of EV existence in the household.
\item The proposed approach does not require manual feature engineering or prior knowledge about charging profiles. 
\item We propose a memory-based transformer that leverage both long and short temporal information to capture different charging pattern, from long charging to slow charging. 
\item The proposed approach uses dual memory compression to ensure that the model achieves linear time complexity and  can run in a real-time manner.
\item We propose to use streaming peak-over-threshold (SPoT) to define threshold value dynamically.
\end{itemize}

The rest of this paper is organized as follows: Section~\ref{sec:proposed_approach} presents the proposed model for online EV charging identification. Section~\ref{Sec:experiments} presents experimental results, and Section~\ref{Sec:conclusion} concludes the paper.

\section{Online Anomaly transformer for EV charging identifications}
\label{sec:proposed_approach}
\begin{figure*}[h]
\centering
\includegraphics[width=0.95\textwidth]{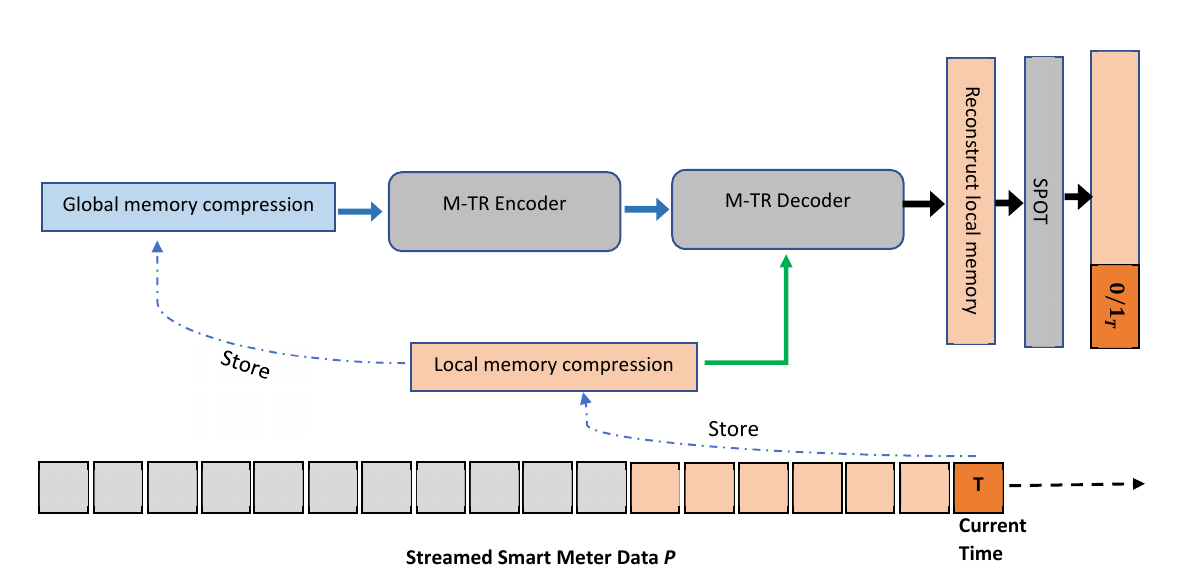}
\caption{Overview of the proposed M-TR for online EV charging identification. Given a live-streaming smart meter readings, M-TR sequentially identifies the EV charging happening at each sampling time $T$ without accessing any future reading. The dashed blue indicates the data flow of memories using the first-in-first-out (FIFO) principle.}
\label{fig:gen}
\end{figure*}

Given the live-streaming smart meter data, the goal is to identify EV charging events in each smart meter sampling time using smart meter readings up to the current time. There is no access to future smart meter readings at the inference phase. Formally, the representation of streaming smart meter readings at time $T$ involves a batch of $B$ past readings denoted as $\mathbf{P}^T=\{P_{T-B+1},\ldots, P_T\}$. The online anomaly identification of EV charging receives $\mathbf{P}^T$ as input and detects the existence of EV charging $\hat{y}_T$,  $\hat{y}_T\in \{0,1\}$. Note that $\hat{y}=0$ reflects no EV charging exists, while $\hat{y}=1$ reflects the existence of EV charging. The overall framework of our model is shown in \Fig~\ref{fig:gen}.
\begin{figure*}
\centering
\includegraphics[width=0.85\textwidth]
{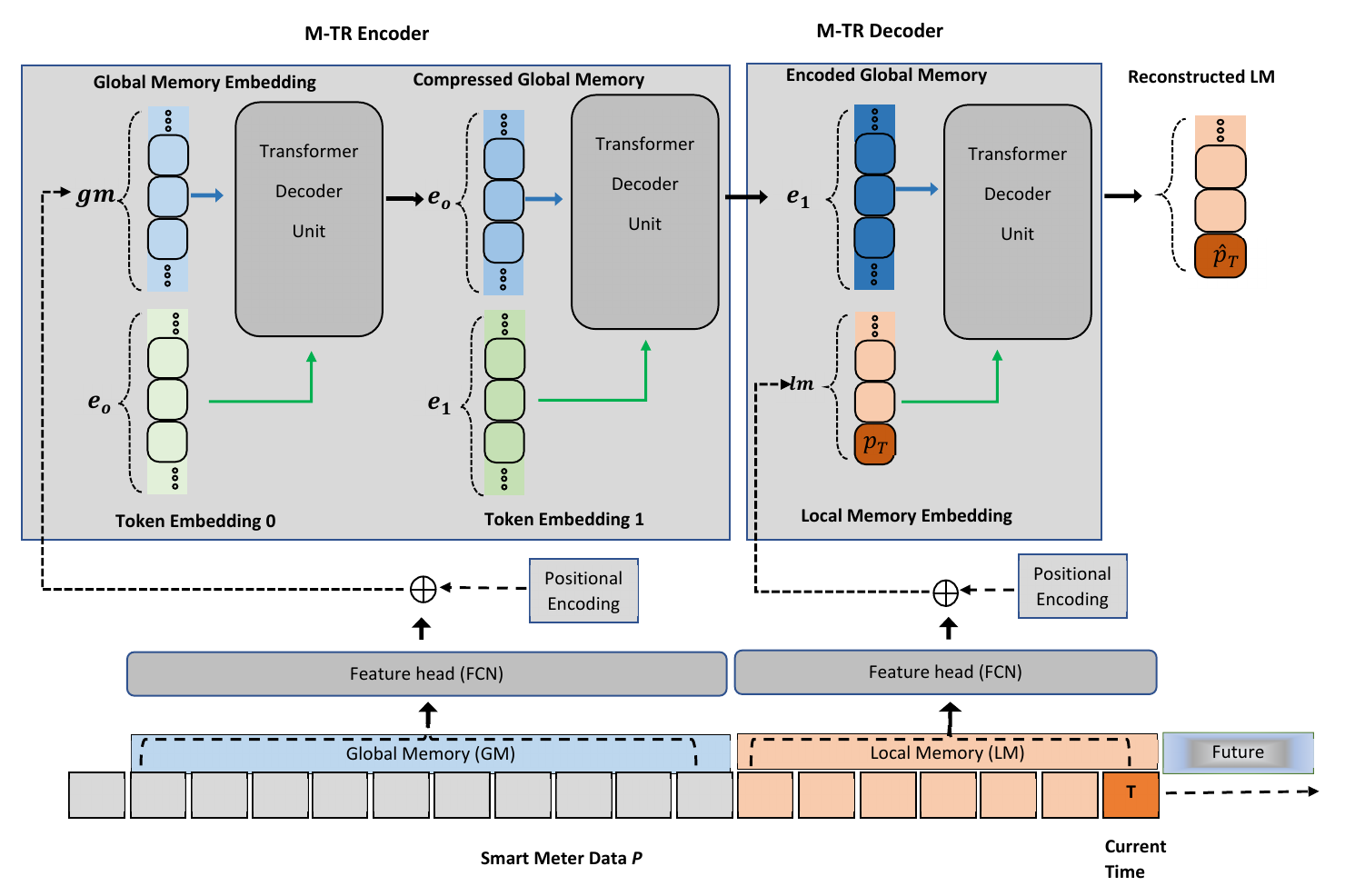}
\caption{Overview of Memory-based Transformer (M-TR) Anomaly Detection (M-TR AD). The network consists of a memory-based transformer (M-TR) encoder and decoder units. M-TR encoder encodes the global memory smart meter reading $gm$ into long memory embedding of size $e_1$ of latent features. The M-TR decoder leverages contextual information from the compressed global memory with the local memory $lm$ for anomaly detection. By referencing relevant context information, the M-TR decoder enhances its ability to detect anomalies in the streamed data. The M-TR encoder and decoder units are based on  the Transformer decoder unit in~\cite{vaswani2017attention}. During the inference phase, M-TR takes every incoming smart meter reading in an online fashion with the absence of future readings. }
\label{fig:MTR}
\end{figure*}
\subsection{Overview}
According to our method, recent smart meter readings provide accurate information about ongoing EV charging events, while long-term smart meter readings provide contextual reference f EV charging that is highly occurring at a particular time. We propose a memory-based Transformer (M-TR), as shown in \Fig~\ref{fig:MTR}.  In our model, we denote the extended smart meter reading  as aglobal memory $gm$ and denote the short-period smart meter readings as a local memory $lm$. The smart meter reading in the extended past ($gm$)  stored  in the global memory embedding, and the local smart meter readings  ($lm$) stores the recent smart meter reading. The M-TR encoder compresses these readings into abstract features as latent features of $e_1$ vectors. The M-TR decoder interacts with both the encoded global memory and local memory in order to decode and reconstruct the local memory readings as $\hat{lm}$. This process involves querying the global memory using the local memory as reference. This design is inspired by using short and long-term memory for action detection in videos~\cite{wu2019long} combined with the advantages of Transformers\cite{vaswani2017attention}. The streamed input smart meter reading is stored in two consecutive memories. In the local memory, only a few recently observed LM readings are stored. In our implementation, we use a first-in-first-out (FIFO) queue with slots of $LM=\{P_T,\ldots,P_{T-sm+1}\}$. The smart meter reading that is older than $lm$ time steps, it enters into the global memory; the global memory $GM$ also works by FIFO principle. The $GM$ can be formatted as $GM=\{P_{T-lm},\ldots,P_{t-lm-gm+1}\}$. Here, the $GM$ works as input memory to M-TR encoder and $LM$ works as queries for M-TR decoder. From a practical point of view, the $gm$ period should be chosen longer than the local memory period $(lm<<gm)$.  The choice of $lm$  and $gm$ is based on the validation set. In our networks, we use a sinusoidal position encoding~\cite{vaswani2017attention} to encode temporal important for each smart meter reading features in local and global memories relative to the current time $T$.

\subsection{M-TR Encoder}
The purpose of M-TR encoder is to encode the global memory $gm$ of meter readings into a compressed feature representations. This representation serves as a valuable resource for that M-TR decoder, enabling it to decode and interpret meaningful temporal context. Capturing the relation and temporal context over a significant number of readings presents a computationally demanding endeavor. To make the training feasible, capturing long temporal context has been previously addressed by either 
temporal sub-sampling~\cite{wu2019long} or using RNNs. However, the previous approaches suffer from losing temporal information at each time step. Recently, Transformers~\cite{vaswani2017attention}, employing an attention-based mechanism, have shown promising results in capturing long temporal modeling. For our M-TR encoder, one example is to choose a self-attention Transformer encoder. However, the self-attention Transformer's time complexity grows quadratically with sequence length $gm$, $O(gm^2C)$, where $C$ is the dimension of smart meter readings features after the FCN. However, a recent work~\cite{wang2020self} demonstrated the utilization of self-attention with liner complexity. This approach involves the repetitive referencing of information from global memory using a multi-layer Transformer architecture. In this paper, we propose to employ  a dual memory comparison for anomaly detection  using Transformer decoder units~\cite{vaswani2017attention} for better memory modeling and encoding.

\textbf{The Transformer decoder unit (TRD)}: The decoder unit employs a Multi-Head Self Attention operation~\cite{vaswani2017attention} with $h$ heads. 
where $\hat{\mathbf{\Lambda}}_i$ is the output of applying self-attention with $Q$ query, $k$ key, and $V$ value matrices, as follows:
\begin{equation}
\hat{\mathbf{\Lambda}}=\text{Attenion(Q,K,V)}=\sigma(\frac{Q\cdot K^T}{\sqrt{C}})V,
\end{equation}
where, in our self-attention implementation, $\mathbf{\Lambda}=Q,K,V$ is the first input tokens~$\mathbf{\Lambda}\in \mathbb{R}^{e\times C}$, $\sigma$ is a softmax operation. The second operation involves applying cross-attention by querying the output of the self-attention, as follows:
\begin{equation}
\text{CrossAttenion}((\delta(\hat{\mathbf{\Lambda}}),\bm{\beta})=\sigma(\frac{\delta(\hat{\mathbf{\Lambda})}\cdot \bm{\beta}^T}{\sqrt{C}})\bm{\beta},
\end{equation}
where $\bm{\beta}$ is $m$ additional input tokens denoted as $\bm{\beta} \in\mathbb{R}^{m\times C}$ which works as both the key and the value matrices in the cross attention operation, $m$ is another large number to be learned, and here $\delta:\mathbb{R}^{e\times C} \to \mathbb{R}^{e\times C}$ acts as an intermediary between the self-attention and the cross attention operations. the advantage of this design lies in its capability to transform input tokens of dimension $m\times C$  into output tokens of the dimension of $e\times C$ with time complexity $O(e^2C+nmC)$.  However, the time complexity becomes linear when $e$ is smaller than $m$, making it particularly suitable for compressing global memory. This technique has proven successful in processing large volumes of input, such as image pixels, as demonstrated in~\cite{jaegle2021perceiver}.

\begin{figure}
\centering
\includegraphics[scale=0.4]{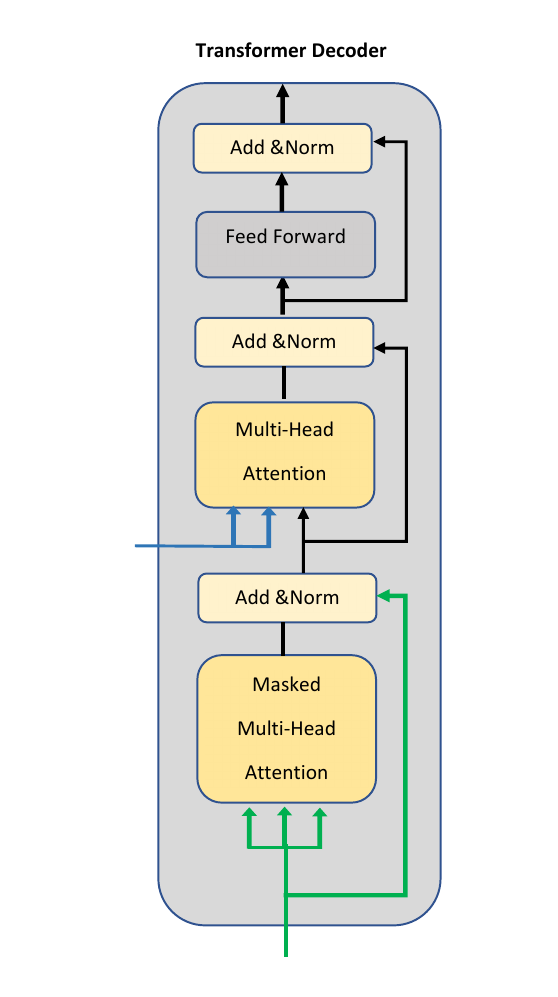}
\caption{Overview of the Transformer decoder unit\cite{vaswani2017attention} used in our M-TR anomaly detection network.}
\label{fig:transformer_decoder_unit}
\end{figure}

\textbf{Dual Memory Compression}: As part of a procedure inspired by~\cite{vaswani2017attention} focused on optimizing global memory, multiple Transformer decoder units are stacked on a global memory. As a consequence, executing the encoder at each time step may require considerable time. An alternative method which addresses this concern is the use of a dual memory compression technique, as shown in in~\Fig ~\ref{fig:MTR}, which results in further reductions in time complexity.

The dual memory compression consists of two stacked TRD units, the input to the first TRD derived from the entire global memory $gm$ and produce tokens with $e_0$ output tokens. Subsequently, the second TRD takes the output of the first TRD unit as inputs and produces a compressed representation of the global memory of size $e_1\times C$. Through the stacking of TRD units, a linear time complexity can be achieved with respect to the global memory size $gm~\times C$. Since $e_0$ and $e_1$ are much smaller than $gm$ and $l_{enc}$, The $O$(M-TR) is reduced to $O(e_0^2C+e_0 gm C+(n_e^2+e_1e_0)l_{enc}C)$.


\subsection{M-TR Decoder}
The local memory carries useful information for identifying whether the EV charging (regarded as anomaly event) is happening at the current time step. The M-TR employs queries to recover valuable information from the compressed global memory by referencing the local memory. The M-TR encoder generates output tokens, which are utilized as input tokens to the M-TR decoder, while the local memory reading $lm$ functions as the output tokens. The M-TR decoder takes these input tokens and produces the reconstructed local memory reading $\hat{LM}$. These readings are used to calculate the anomaly scores  $As_T$, which will be discussed in the next section.

\subsection{Anomaly score}
\label{sec:anomaly_score}
This section explains the general framework for identifying EV charging. Given short-memory smart meter readings $LM$,  with  length $lm$, We define the Mean Squared Error (MSE) as the anomaly score $A_s$ between the short-memory smart readings and the reconstructed ones as follows:

\begin{equation}
A_s=\sum_{lm}(LM_i-\hat{LM}_i)/N,
\end{equation}
where $\hat{LM}_i$ is the reconstructed smart meter readings obtained from our model, and $N$ is the total number of observations that can be defined based on the batch size during the training phase. 
After obtaining the anomaly score $As$, if this anomaly score is above the specified threshold  $thre$ $As_T> thre$, then $As_i=1$ indicates an anomaly. In this work, we choose $thre$ dynamically by using Streaming Peak-Over-Threshold (SPOT).
The general algorithm of our proposed model is shown in Algorithm~\ref{alg:1}. The input to our algorithm consists of several components. Firstly, we have the smart meter reading, which provides information about the energy consumption up to a specific time, denoted as $T$. second, the MTR  model with pretrained weights, represented by $\theta$. Furthermore, the algorithm takes into account the length of both local memory and global memory.  The output of the algorithm is a set of charging indicators, which also serve as anomaly indicators. These indicators provide insights into the charging behavior at the specific time $T$. 
\begin{algorithm}[t]
\begin{algorithmic}[1]
	\Procedure{Anomaly\_Detection}{$\mathbold{P}_T; \text{M-TR}(;\theta); gm; lm$}
	\Comment \textcolor{\commentcolor}{$\mathbold{P}_T:$ smart meter power consumption data of length $t-lm-gm+1$; $\text{M-TR}(;\theta):$ M-TR network with pretrained parameters $\theta$; $lm$: local memory length; $gm$: global memory length}   
	\State $P_T=\{LM=\{P_T,\dots, P_{T-lm+1}\},GM=\{P_{T-lm},\dots,P_{t-lm-gm+1}\}\}\gets$ \Call{Process sequence}{$\mathbold{P}_T$}
	\Comment \textcolor{\commentcolor}{Use FIFO to generate local and global memory segments.}
	\State $y_T \gets \phi$
	\Comment \textcolor{\commentcolor}{Initialize the anomaly score with $\phi$}
	\State $\hat{LM}\gets$ \Call{M-TR}{$LM;GM;\theta$}
	\Comment \textcolor{\commentcolor}{Forward pass to generate the reconstructed local memory conditioned on the global memory}
	\State $As \gets$ \Call{MSE}{$ LM,\hat{LM}$}
	\Comment \textcolor{\commentcolor}{Calculate the reconstruction error} 
	\State $A_{th} \gets$ \Call{SPoT}{$As$}
	\Comment \textcolor{\commentcolor}{Calculate the anomaly threshold using streaming-peak-over-threshold }
	\If{$As_T > {A}_{th}$} 
	\State $y_T \gets 1$ 
	\Comment \textcolor{\commentcolor}{EV charging detected.}
	\Else
	\State $y_T \gets 0$ 
	\EndIf
	\State \textbf{return} $y_T$
	\Comment \textcolor{\commentcolor}{EV charging  not detected.}
	
	\EndProcedure	
\end{algorithmic}  
\caption{M-TR online EV charging detection.\label{alg:1}}
\end{algorithm}
\subsection{Streaming peak-over-threshold for dynamic anomaly threshold selection}
\label{sec:spot}
In the Extreme Value Theory (EVT), the peak-over-threshold method (PoT) is a popular statistical method for quantifying the likelihood of rare (extreme) observations of a variable. Let $X$ to be a streaming time series of independent and identical distributed samples.  Pickands-Balkema-de Haan theorem~\cite{pickands1975statistical} (also referred to as the second theorem in EVT in comparison to Fisher, Tippett and Gnedenko's initial result) is the basis of PoT as shown below:

\begin{equation}
\textbf{F}_h(x)=\mathbb{P}(X-h>x|X>h) \underset{h \to \tau}{\sim} (1+\frac{\gamma x}{\sigma(h)})^{\frac{-1}{\gamma}},
\end{equation}
where $\textbf{F}$ is the accumulative distribution function $\textbf{F}\in \mathcal{D}_\gamma$ if and only if a function $\sigma$ exists for all $x\in \mathbb{R}~s.t 1+\gamma x>0$.
PoT uses the Generalized Pareto Distribution (GPD) with parameters $\gamma, \sigma$ to find the excess over a threshold $h$, written $X-h$. In this case, we try to estimate the parameter $\hat{\gamma}$ and $\hat{\sigma}$. Different classical methods can be used to calculate these parameters, such as the Method of Moments or the Probability Weighted Moments, but in this paper, we used Maximum Likelihood Estimator (MLE) method ~\cite{siffer2017anomaly} due to its proven efficiency~\cite{tuli2022tranad}.  In practice, the log-likelihood $\mathcal{L}$ is used to estimate the parameters that maximize GPD distribution.

\begin{equation}
\hat{\gamma},\hat{\sigma}= \argmax_{\gamma,\sigma}\log{\text{GPD}}.
\end{equation}
For the GPD distribution, the MLE turns into:
\begin{equation}
log \mathcal{L}(\sigma, \gamma)=-N_h\log\sigma -(1+\frac{1}{\gamma})\sum_{i=1}^{N_h}\log(1+\frac{\gamma}{\sigma}Y_i),
\end{equation}
where $Y_i> 0$ are the excesses of $X_i$ over $h$.
The quantile $z_q$ of the distribution can be computed as follows:
\begin{equation}
z_q=h+\frac{\hat{\sigma}}{\hat{\gamma}}((\frac{qn}{N_h})^{-\hat{\gamma}}-1),
\end{equation}
where $q$ is the desired probability, $n$ is the total number of observations, $N_h$ is the number of peaks. In this paper, we use streaming peak-over-threshold (SPoT) as  described 
in Algorithm~\ref{alg:1}. SPoT uses PoT to obtain a baseline threshold ($z$) and an initial threshold ($h$). The threshold will be updated during this process of sequentially traversing all the test data points.

In SPoT, $A_{si}$ can be classified into three types: Normal : $A_{si}$ below the initial threshold; Peaks: when $A_{si}$ is between the initial threshold and base threshold, such datapoints will not be detected as an anomaly but will affect the threshold; Anomaly: when $A_{si}$ is over the base threshold, such datapoints will be detected as anomaly datapoints directly, will not affect the threshold. For generality, we use $X$ instead of $As$ in the Algorithm~\ref{alg:2}.

\begin{algorithm}[t]
\begin{algorithmic}[1]
	\Procedure{SPoT}{$(X_i)_{i>0}, n,q$}
	\Comment $:\mathbold{X}_i$ Reconstructed anomaly scores generated by M-TR
	\State $\textbf{AT} \gets \phi$
	\Comment \textcolor{\commentcolor}{Initialize the anomaly score }
	\State $z_q,h \gets$ PoT $(X_1\ldots, X_n,q)$
	\State $k \gets n$
	\For{$i > n$}
	\If{$X_i >z_q$} 
	\Comment{\textcolor{\commentcolor}{Anomaly case (EV charging)}}
	\State $\text{Add} (i,X_i)~\text{in}~\textbf{AT}$ 
	\ElsIf{$X_i >h$}
	\State $Y_i \gets X_i-h $
	\State Add $Y_i$ in $\mathbf{Y_t}$
	\State $N_h \gets N_h+1$
	\State $k \gets k+1$
	\State $ \hat{\sigma},\hat{\gamma} \gets$ GRIMSHAW $(\mathbf{Y}_t)$
	\State $z_q \gets $ CalcThreshold $(q,\hat{\sigma},\hat{\gamma},k,N_h,h)$
	\Else 
	\Comment{\textcolor{\commentcolor}{Normal}}
	\State $k \gets k+1$
	
	\EndIf
	\EndFor
	\State \textbf{return} $\eta$
	\EndProcedure	
\end{algorithmic} 
\caption{Streaming PoT (SPoT).\label{alg:2}}
\end{algorithm}

\subsection{Online Inference with M-TR}
During the online inference, as time progresses, the streamed smart meter readings are continuously fed to the M-TR model. However, executing the M-TR global memory encoder from scratch for each new smart meter sampling recordings leads to time complexity of $O(e_0^2C+e_0gmC)$, while for the second memory stage it turns into $O((e_1^2+e_1e_0)l_{enc}C)$. Nonetheless, it is important to mention that at each time step, only one smart meter reading needs to be updated. To improve the efficiency of online inference, storing intermediate results of the M-TR decoder can be implemented. We can fix the queries of the M-TR decoder, thus the pre-computed self-attention outputs can bed used throughout the inference. Thus, the cross attention of M-TR encoder for relative position $\tau=T-t$ of reading $P$ at time $t$ can be expressed as follows:

\begin{dmath}
\text{CrossAttention}(\mathbf{q}_i,{P_{T-\tau}+s_{\tau}})= \\ \sum_{\tau=lm}^{lm+gm-1} \frac{\mathcal{Z}}{\sum_{\tau=lm}^{lm+gm-1}\mathcal{Z}} \cdot (P_{T-\tau}+s_{\tau}),\\
\end{dmath}
where $\mathcal{Z}$ is defined as follows:
\begin{equation}
\mathcal{Z}= \exp((P_{T-\tau}+s_{\tau}) \cdot \mathbf{q}_i \/ \sqrt{C}).
\end{equation}
The computation we discussed relies on the attention weight matrix $\mathbf{A}$, which has dimensions $\mathbb{R}^{gm \times e_0}$. Each element $a_{\tau i}$ of this matrix is determined by the product of $P_{T-\tau +s_{\tau}}$ and $\mathbf{q}i$. To further analyze $\mathbf{A}$, we can decompose it into two distinct matrices, namely $\mathbf{A}^p$ and $\mathbf{A}^s$. These matrices consist of elements $a{\tau i}^p=P_{T-\tau} \cdot \mathbf{q}i$ and $a{\tau i}^s=s_{\tau} \cdot \mathbf{q}_i$, respectively. It is important to note that during the inference phase, both the queries following the initial self-attention step and the position embedding $s_{\tau}$ remain unchanged. This results in pre-use of matrix $\mathbf{A}^s$ for every meter readings update. Moreover, $\mathbf{A}^p$ can be computed at each time step $T$ by stacking all vectors using FIFO queue $a_t=\mathcal{\mathbf{Q}}^{'} P_t$ of size $gm$. This procedure has a time complexity of $O(e_0C)$. We can see that matrix $\mathbf{A}$ can be computed with only $e_0\times gm$ instead of $e_0 \times gm \times C$.

\section{Experiments}
\label{Sec:experiments}
\subsection{Dataset}
\begin{figure*}
\centering
\subfloat[User 1]{\includegraphics[width=0.75\textwidth]{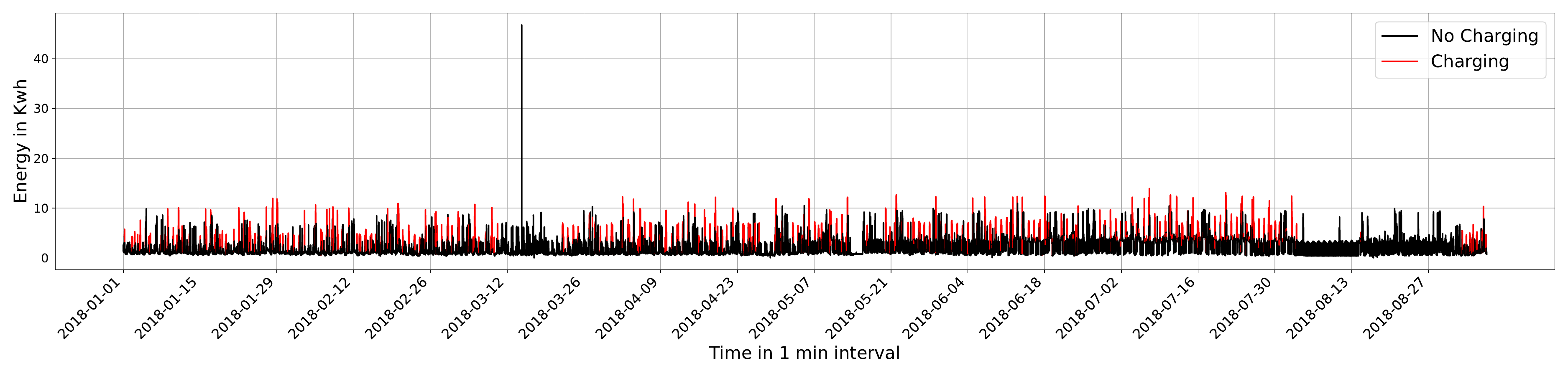}}\\
\subfloat[User 3]{\includegraphics[width=0.75\textwidth]{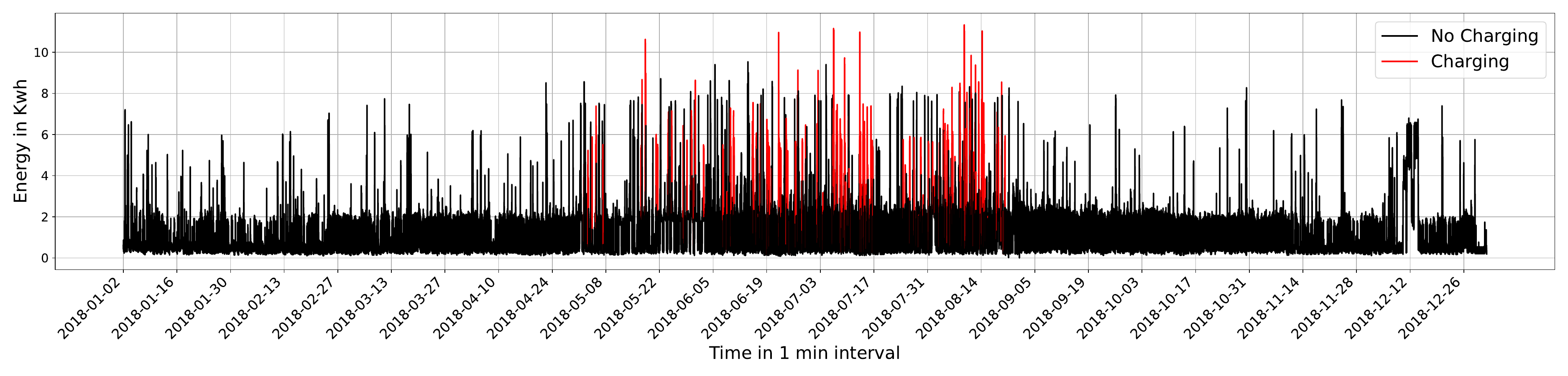}}\\
\subfloat[User 4]{\includegraphics[width=0.75\textwidth]{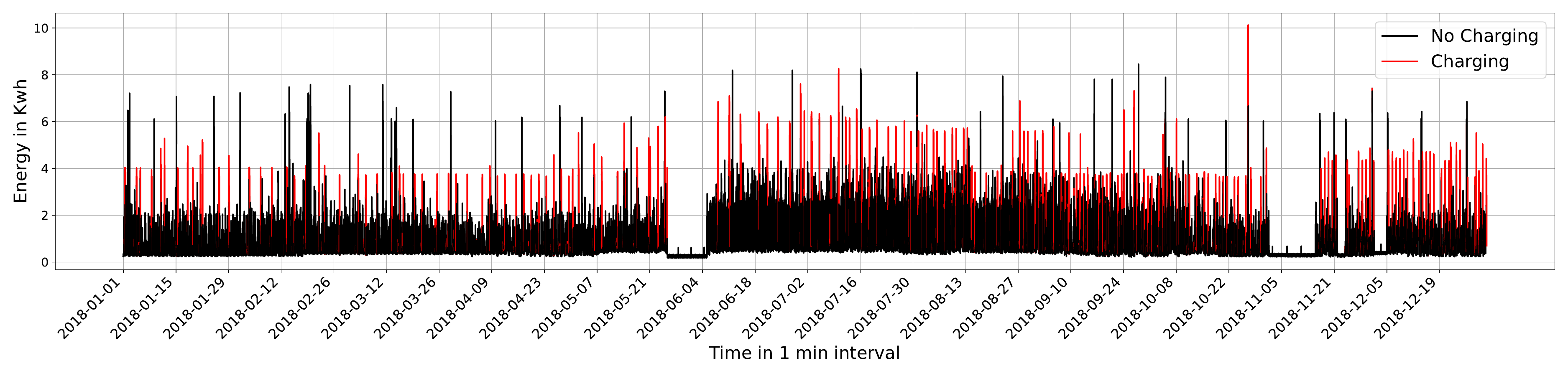}}\\
\subfloat[User 5]{\includegraphics[width=0.75\textwidth]{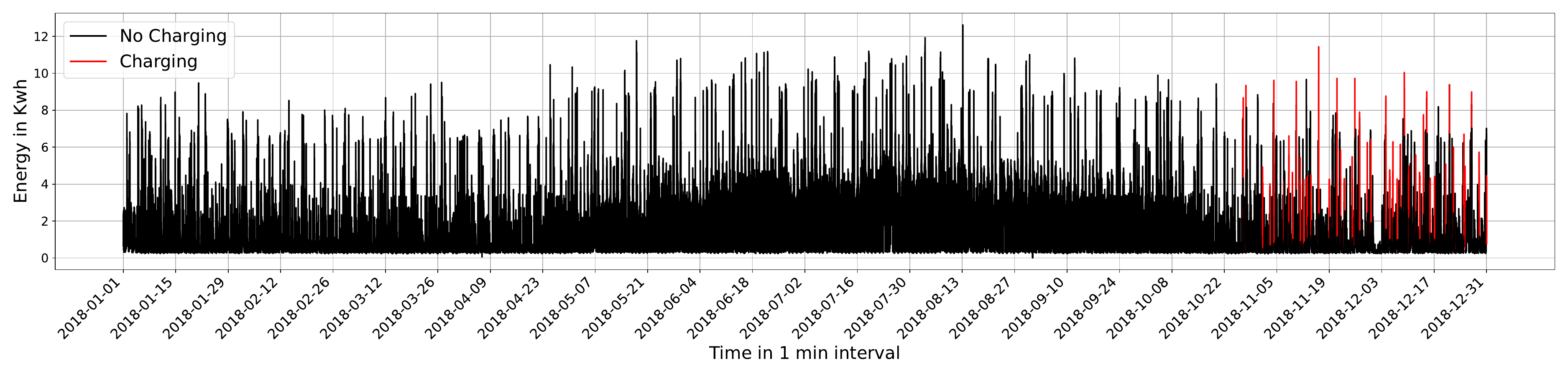}}\\
\subfloat[User 6]{\includegraphics[width=0.75\textwidth]{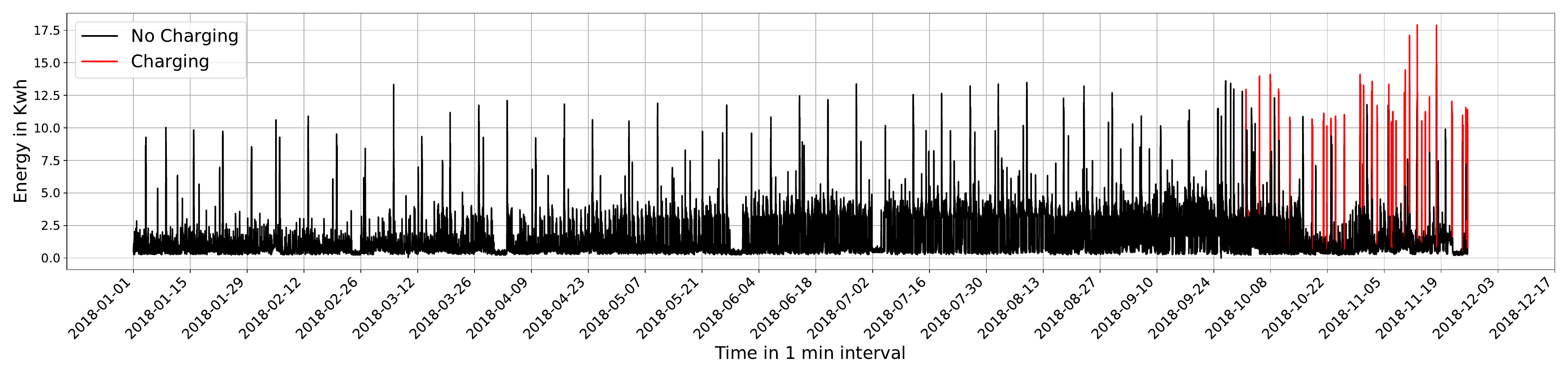}}\\
\subfloat[User 8]{\includegraphics[width=0.75\textwidth]{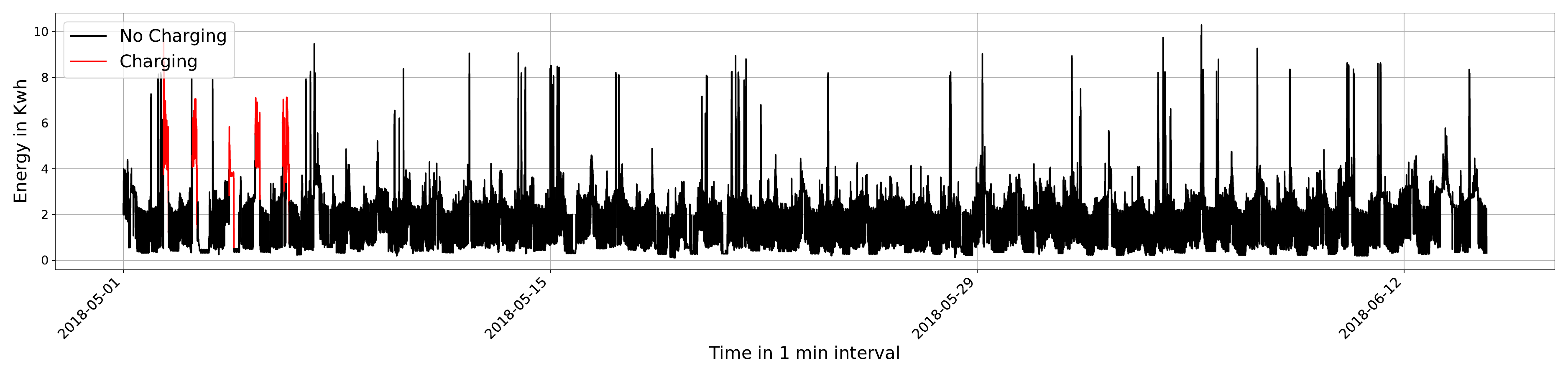}}\\
\caption{Examples of the smart meter recording in the dataset used in our experiments and their anomaly distribution (EV-charging) as shown in red color.}
\label{fig:dataset_samples}
\end{figure*}

\begin{figure*}
\centering
\subfloat[User 1]{\includegraphics[width=0.7\textwidth]{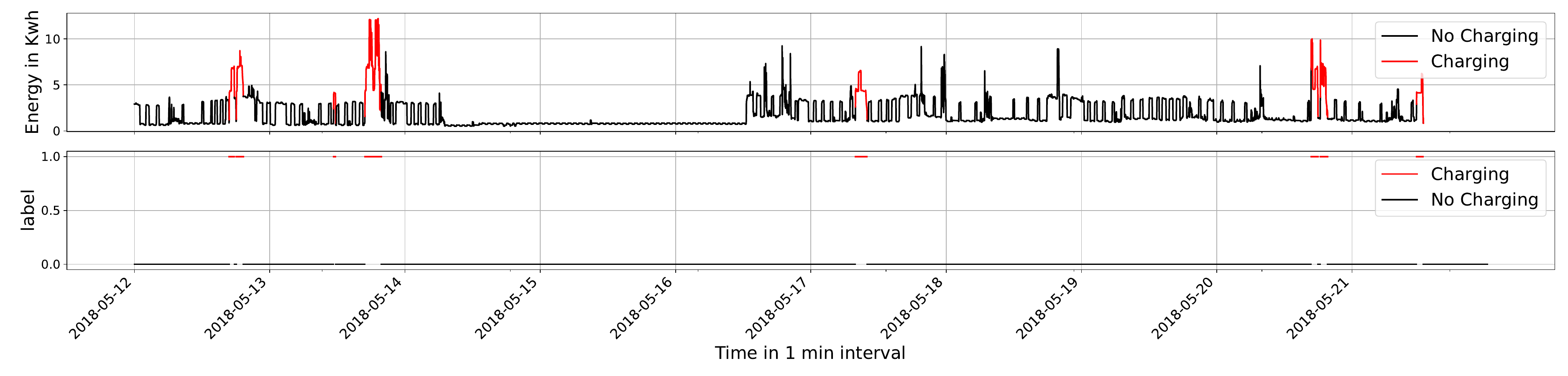}}\\
\subfloat[User 3]{\includegraphics[width=0.7\textwidth]{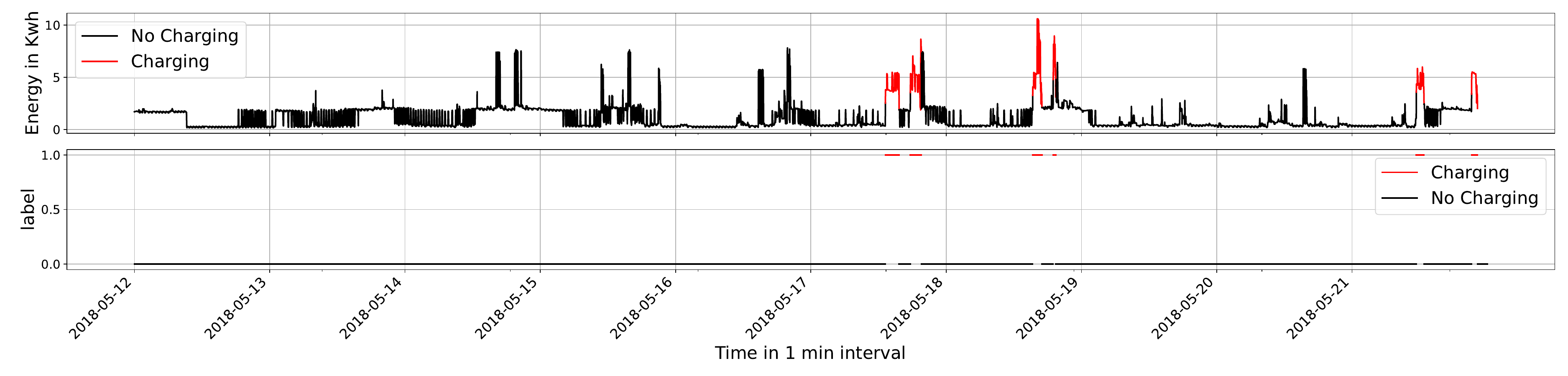}}\\
\subfloat[User 4]{\includegraphics[width=0.7\textwidth]{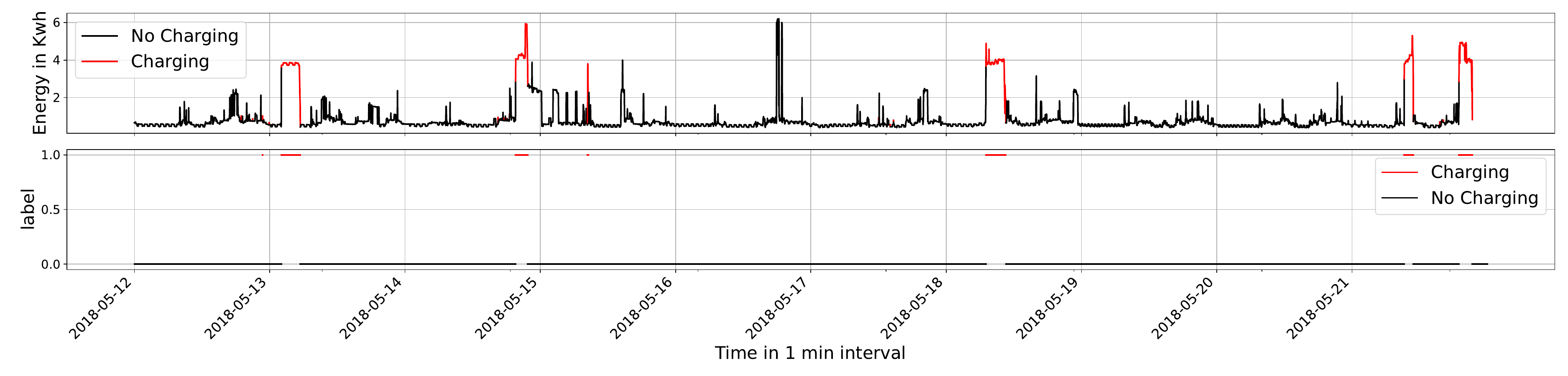}}\\
\subfloat[User 5]{\includegraphics[width=0.7\textwidth]{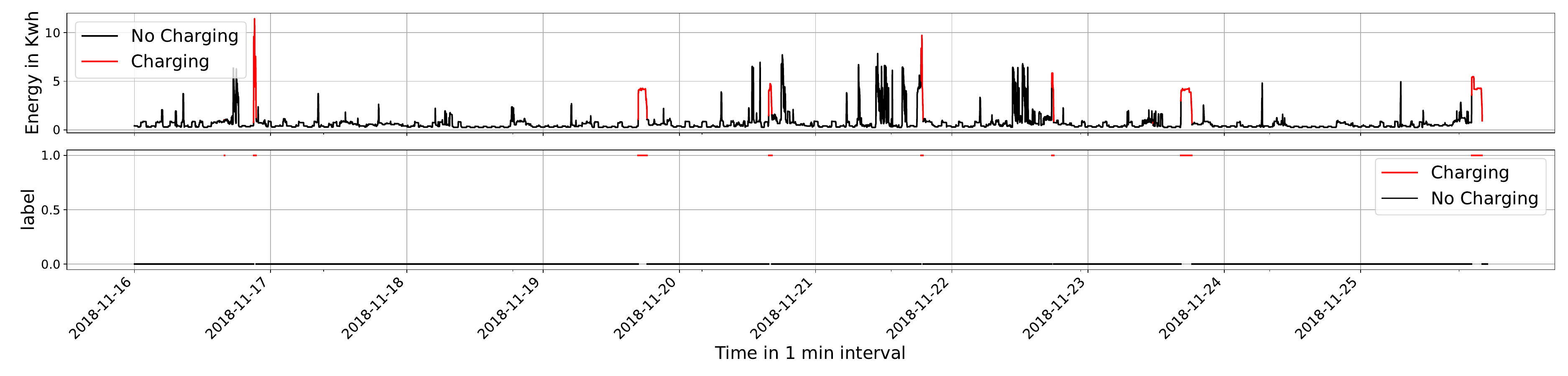}}\\
\subfloat[User 6]{\includegraphics[width=0.7\textwidth]{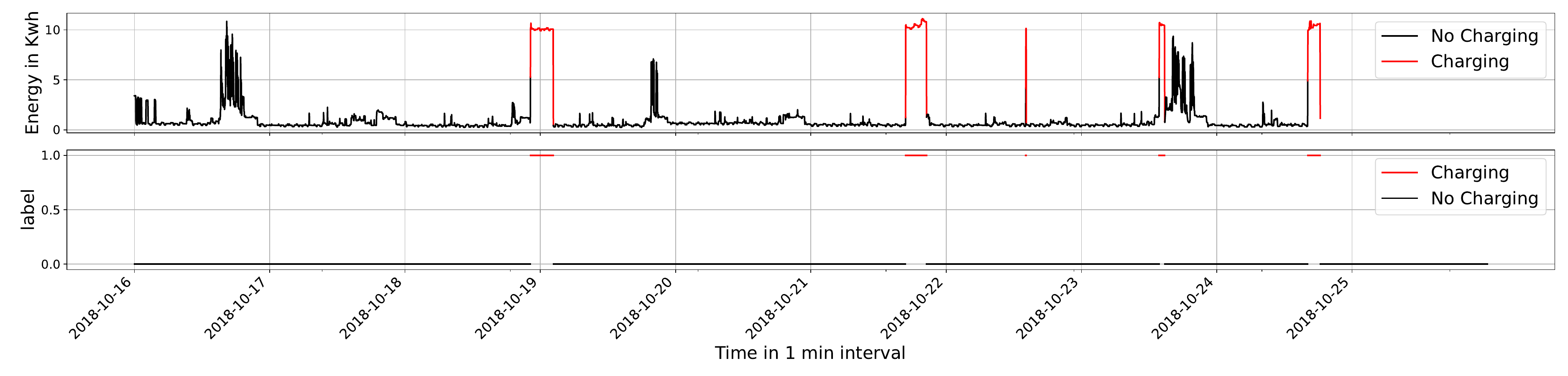}}\\
\subfloat[User 8]{\includegraphics[width=0.7\textwidth]{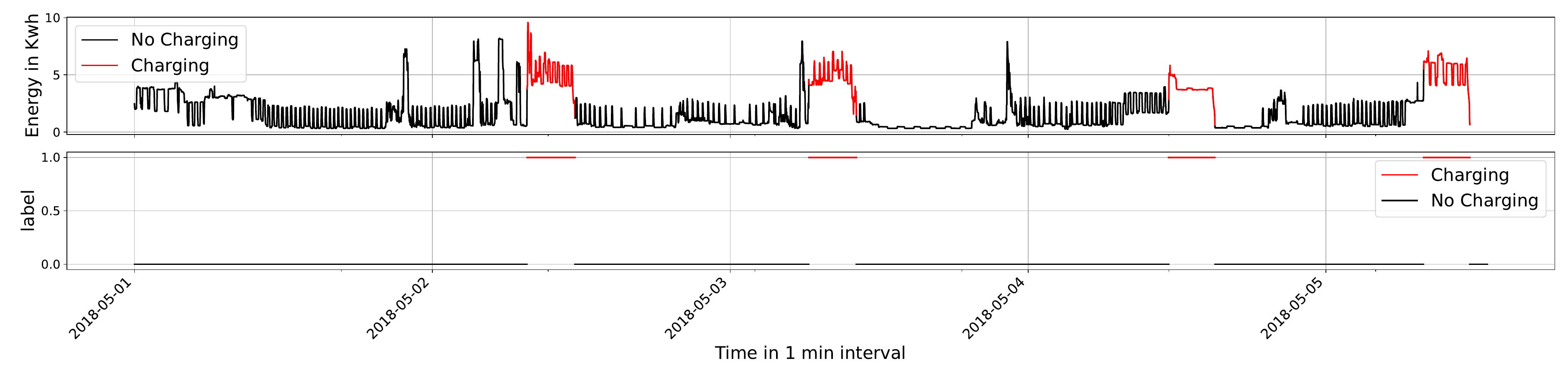}}\\
\caption{Snapshot of the smart meter recordings in the dataset used in our experiments and their temporal annotations of EV charging label as shown in the red color.}
\label{fig:dataset_samples_snap}
\end{figure*}

The goal  of this paper is to run online unsupervised anomaly detection for EV charging identification.  We used Pecan Street Dataport~\cite{street2019pecan}, which provides a set of labeled EV charging events for 8 households. However, we do not use these labels during the training phase. The Pecan Street dataset provides minute-level of household energy consumption with labeled EV charging levels. Geographically, we used data obtained from the Pecan Street dataset with minute-level residential load data with EV charging load in Austin, Texas, over one year (January 1, 2018, to December 31, 2018). \Fig~\ref{fig:dataset_samples} shows examples of the entire time series data of five users, namely users 1, 3, 4, 5, 6,7, and 8. This dataset has different EV charging distribution over the entire readings, some users charge their EV for a specific period and then stop charging, such as user 3. Capturing such behavior is also of interest. \Fig~\ref{fig:dataset_samples_snap} shows a snapshot of these users and the temporal annotation of their charging for a period of six days. We can observe that these users have different EV charging intervals.\\ 

\begin{figure*}
\centering
\includegraphics[width=0.8\textwidth]{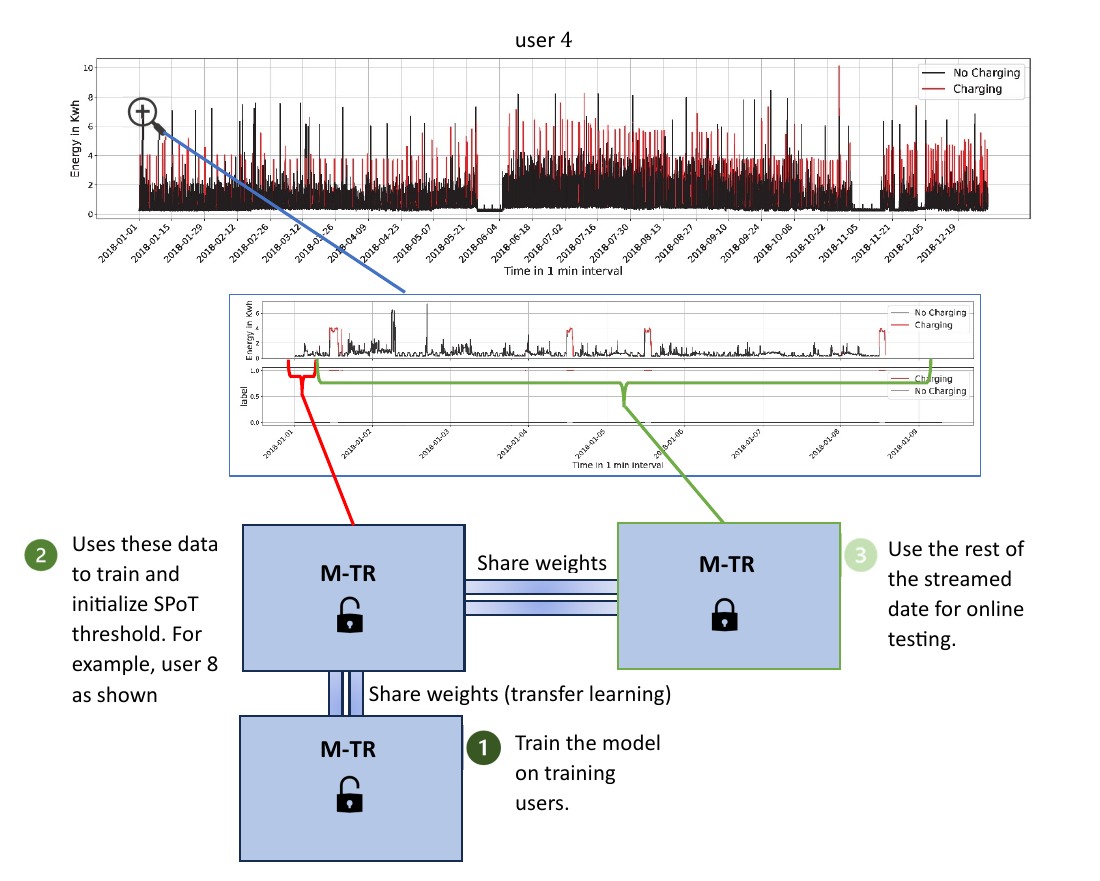}
\caption{Online testing phase of our M-TR for user 4. This procedure is used for all testing users.}
\label{fig:testing_phase}
\end{figure*}

In our experiments, we need to train the model on non-EV charging data, thus, we chose the non-EV charging intervals of users 2, 6, and user 8 for training and transfer the learning into the other users for testing. The main reason for choosing these users is because these users have few charging intervals, and a lot of non-Ev charging intervals as shown in~\Fig~\ref{fig:dataset_samples}. Since our M-TR model is an online model, during the test phase, we initialize our model using the starting non-EV charging interval to adjust the threshold value of SPOT discussed in Section~\ref{sec:spot} as shown in~\ref{fig:testing_phase}. We chose a maximum of 4 weeks of non-EV charging interval to train our model. if no such data is available, we use the available non-EV intervals to train the model. 
The statistics of these users and the percentage of training size and test size are shown in Table~\ref{dataset_stat}. Due to the dataset nature, there are different training sizes (when no EV charging is happening) that can be used to train the model for each user. 


\begin{table*}[htbp]
\centering
\scriptsize
\caption{Statistics of the testset used in our experiment.}
\begin{tabular}{lrrrrrr}
	\toprule
	User ID & Train start (normal) (M/d/y) &Train end& Train size \& percentage \% & Test start & Test end & Test size \\
	\toprule
	User 1 & 1/1/2018 11:00 & 1/1/2018 15:59 & 300 (0.08\%) & 1/1/2018 16:00 & 12/9/2018 9:59 & 357709 \\
	User 3 & 1/2/2018 3:00 & 1/30/2018 08:59 & 40320   (7.86\%)  & 1/30/2018 09:00 & 12/04/2018 11:00 &  472297 (92\%)\\
	User 4 & 1/1/2018 11:00 & 1/1/2018 23:59 & 174 (0.033\%) & 1/1/2018 13:54 & 1/1/2019 10:59 & 522146 \\
	User 5 & 1/1/2018 11:00 & 01/29/2018 16:59 & 40320 (7.6\%) & 01/29/2018 17:00 & 12/31/2018 19:32 & 484806 (92\%) \\
	\\
	\bottomrule
\end{tabular}%
\label{dataset_stat}%
\end{table*}%

\subsection{Evaluation protocol}
We consider an average precision, recall, and $F_1$ score as metrics for the performance evaluation. We also include the Area Under the Receiver Operator Characteristic (ROC), which measures the area under the true positive rate as a function of the false positive rate. Precision and recall are defined as,
\begin{equation}
\text{Precision}=\frac{TP}{TP+FP}, \text{Recall}=\frac{TP}{TP+FN},
\end{equation}
where $TP$ is the true positive rate, $FP$ is the false positive rate, and $FN$ is the false negative rate. The $F_1$ is expressed as,
\begin{equation}
F_1=2\times\frac{\text{Precision}\times \text{Recall}}{\text{Precision}+\text{Recall}}.
\end{equation}

\subsection{Baseline methods}
This section explores different state-of-art anomaly detection models used to benchmark against our method.
\textbf{LTSM-NDT}~\cite{hundman2018detecting} (with autoencoder implementation from openGauss~\cite{li2021opengauss}: This model uses an LTSM-based deep neural model that divides the time series data into input sequences and uses a forecasting approach to predict the next timestamp. LTSM relies on the autoregressive approach that learns temporal dependency. LTSM-NDT uses non-parametric dynamic error thresholding (NDT) to set a threshold for labeling anomalies.

Deep autoencoder with Gaussian Mixtures Models (\textbf{DAGMM})~\cite{zong2018deep}: The distribution assumption of this model has been imposed on the latent space. DAGMM imposes Gaussian mixture distribution on the latent distribution. DAGMM uses a combination of the reconstruction loss, sample energy, and GMM covariance to train the model. In DAGMM,  the sample energy is used as an anomaly score. We use the Pytorch implementation of the model on GitHub\footnote{https://github.com/mperezcarrasco/PyTorch-DAGMM}.

The Multi-Scale Convectional Recursive Encoder-Decoder (\textbf{MSCRED})~\cite{zhang2019deep}: MSCRED uses a sliding window approach to convert the time series data into a normalized two-dimensional image and then is fed into the ConvLTSM layer. This network is capable of capturing complex inter-modal correlations and also temporal dependencies. While this method captures more complex correlations between modalities and temporal information, it cannot be generalized in settings without adequate training data.

\textbf{MTAD_GAT}~\cite{zhao2020multivariate}: This model uses a graph-attention-based network to learn features and temporal correlations and is then fed into Gated Recurrent Unit (GRU) to detect anomalies. The attention mechanism performs compression using a convex combination where the model weights are determined using neural networks. GRU offers a simplified version of LTSM with fewer parameters and can be trained with limited data.

\textbf{CAE-M}~\cite{zhang2021unsupervised}: This model uses a convolutional autoencoder-based network with memory. This network passes time series data into convolutional networks and then fed to Bidirectional LTSM to capture long-term temporal dependencies. These models have been shown to have high computational costs and can not be used for online inference.  

\textbf{USAD}~\cite{audibert2020usad}: This technique uses autoencoder-based architecture with an adversarial-based game style for training the model. This network uses fewer computational resources than the attention-based network's recurrent models.

\textbf{GDN}~\cite{deng2021graph} (with graph embedding implementation from GraphAn~\cite{boniol2020graphan}): GDN uses a graph network to learn the relationship between data modes by using graphs combined with attention-based network. Anomaly scores are calculated based on deviation scores. We also compare our model with a recent paper that uses a transformer network for anomaly detection. 
\textbf{TranAD}~\cite{tuli2022tranad}: This model uses a transformer-based network that allows single-shot inference using time series data using positional encoding. This model is proposed for multivariate time series data. It uses self-conditioning adversarial training to enhance training stability. Despite this model showing less training time, it still takes longer due to its complexity. Similar to our work, this model uses Peak-Over-Threshold to define threshold dynamically. 

For all baseline models, we use the hyperparameters as presented in their papers. All models were implemented using PyTorch 1.7.1 library. For our model, we use the Adam optimizer~\cite{kinga2015method} to train the model with weight decay of $5 \times 10^{-5}$, and base learning rate of $7\times 10^{-5}$ with the momentum of $0.9$. For transformer units, we set the number of heads to 2 and hidden units to 8. The size of short memory is set to 8 and the size of long memory is set to 32.

\subsection{Experimental results}
This section explores the proposed approach's performance against the baseline methods.
Table~\ref{Tab:results} provides the precision, recall, F1, and ROC scores for different models, including our model (M-TR). The proposed online M-TR outperforms all models for four users except for user 3, where CAE-M performs better. On average, the mean of the F1 score is $84\%$, and the mean of ROC is $91\%$ as shown in Table~\ref{Tab:mean_results}. Our model outperforms the transformer-based model (TranAD) for all users, which demonstrates the effectiveness of our memory-based transformer model.  The LTSM-NDT performs worse on all users with a mean F1 score of $42.9\%$ due to its sensitivity and poor efficiency of the NDT thresholding method~\cite{zhao2020multivariate}. DAGGM performs much worse for most users and shows better performance for only user 4 when the training samples are very short which is due to incapability in mapping the long temporal sequence because it does not use window sequence. The MSCRED and CAE-M use sequential data as input that preserves the temporal dependencies, and they also use reconstruction error which prevents them from detecting anomalies closer to normal trend~\cite{audibert2020usad}. The CAE-M shows only better performance in F1 score for user 3. In contrast, MSCRED performs better than other baseline models. USAD, MTAD-GAT, and GDN use attention-based technique that focuses on specific data modes. Despite these models trying to capture long-term trends using local windows, they still perform worse for all users.

The M-TR online inference execution time is 1.2 sec. for each 1-minute reading, which is adequate. The experiments were conducted using the following PC specifications: an AMD Ryzen 5 4600H processor and 16 GB of RAM.

\begin{table*}
\caption{Performance comparison of M-TR with baseline methods on the testset. Pre.: Precision, Rec.: Recall, ROC: Area under the ROC curve, F1: F1 score. The best F1 and ROC scores are highlighted in bold.}
\label{Tab:results}
\scriptsize
\begin{tabular}{llllllllllllllllllll}
	\toprule
	\multicolumn{5}{c}{\textbf{User  1}} &  & \multicolumn{4}{c}{\textbf{User 3}} &  & \multicolumn{4}{c}{\textbf{User 4}} &  & \multicolumn{4}{c}{\textbf{User 5}} \\
	\toprule
	\textbf{Model} & \multicolumn{1}{c}{{ \textbf{F1}}} & \multicolumn{1}{c}{{\textbf{Pre.}}} & \multicolumn{1}{c}{{ \textbf{Rec.}}} & \multicolumn{1}{c}{{\textbf{ROC}}} &  & \multicolumn{1}{c}{{ \textbf{F1}}} & \multicolumn{1}{c}{{ \textbf{Pre.}}} & \multicolumn{1}{c}{{\textbf{Rec.}}} & \multicolumn{1}{c}{{ \textbf{ROC}}} &  & \multicolumn{1}{c}{{ \textbf{F1}}} & \multicolumn{1}{c}{{\textbf{Pre.}}} & \multicolumn{1}{c}{{ \textbf{Rec.}}} & \multicolumn{1}{c}{{ \textbf{ROC}}} &  & \multicolumn{1}{c}{{ \textbf{F1}}} & \multicolumn{1}{c}{{ \textbf{Pre.}}} & \multicolumn{1}{c}{{ \textbf{Rec.}}} & \multicolumn{1}{c}{{ \textbf{ROC}}} \\
	& - & - & - & - &  &  &  &  &  &  &  &  &  &  &  &  &  &  &  \\
	LTSM-NDT & 0.368 & 0.225 & 0.997 & 0.844 &  & 0.449 & 0.777 & 0.315 & 0.654 &  & 0.479 & 0.315 & 1.000 & 0.608 &  & - & - & - & - \\
	DAGMM & 0.308 & 0.332 & 0.286 & 0.617 &  & 0.313 & 0.830 & 0.193 & 0.595 &  & 0.741 & 0.805 & 0.687 & 0.813 &  & - & - & - & - \\
	USAD & 0.045 & 0.060 & 0.036 & 0.492 &  & 0.327 & 0.824 & 0.204 & 0.600 &  & 0.529 & 0.422 & 0.113 & 0.529 &  & 0.028 & 0.812 & 0.014 & 0.507 \\
	MSCRED & 0.647 & 0.479 & 0.997 & \cellcolor[HTML]{8EA9DB}\textbf{0.949} &  & 0.813 & 0.897 & 0.744 & 0.869 &  & 0.536 & 0.367 & 0.997 & 0.688 &  & - & - & - & - \\
	CAE\_M & 0.409 & 0.408 & 0.411 & 0.678 &  & \cellcolor[HTML]{8EA9DB}\textbf{0.935} & 0.927 & 0.944 & 0.969 &  & 0.807 & 0.822 & 0.792 & 0.865 &  & 0.009 & 1.000 & 0.005 & 0.502 \\
	GDN & 0.041 & 0.054 & 0.033 & 0.491 &  & 0.172 & 0.412 & 0.108 & 0.526 &  & 0.172 & 0.412 & 0.108 & 0.526 &  & 0.056 & 0.695 & 0.029 & 0.514 \\
	MTAD\_GAT & 0.562 & 0.506 & 0.631 & 0.788 &  & 0.735 & 0.870 & 0.637 & 0.814 &  & 0.852 & 0.834 & 0.870 & 0.904 &  & 0.009 & 1.000 & 0.005 & 0.502 \\
	TranAD & 0.124 & 0.073 & 0.408 & 0.469 &  & 0.728 & 0.870 & 0.626 & 0.809 &  & 0.248 & 0.529 & 0.162 & 0.555 &  & 0.420 & 0.934 & 0.271 & 0.635 \\
	\textbf{Online M-TR} & \cellcolor[HTML]{8EA9DB}\textbf{0.874} & 0.865 & 0.883 & 0.935 &  & 0.904 & 0.859 & 0.955 & \cellcolor[HTML]{8EA9DB}\textbf{ 0.971} &  & \cellcolor[HTML]{8EA9DB}\textbf{0.917} & 0.913 & 0.920 & \cellcolor[HTML]{8EA9DB}\textbf{0.944} &  & \cellcolor[HTML]{8EA9DB}\textbf{0.753} & 0.829 & 0.690 & \cellcolor[HTML]{8EA9DB}\textbf{0.840}\\
	\toprule
\end{tabular}
\end{table*}

\begin{figure*}
\centering
\subfloat[User 1]{\includegraphics[width=0.85\textwidth]{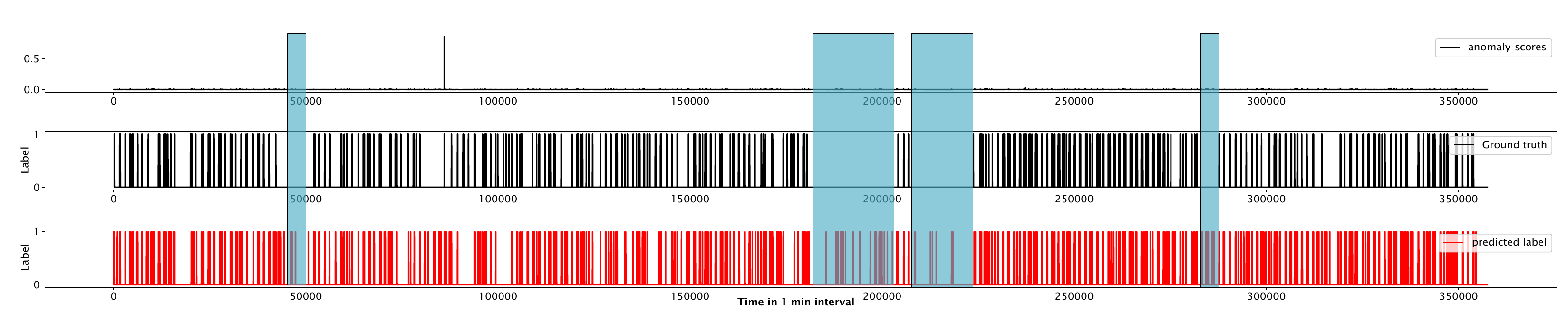}}\\
\subfloat[User 3]{\includegraphics[width=0.85\textwidth]{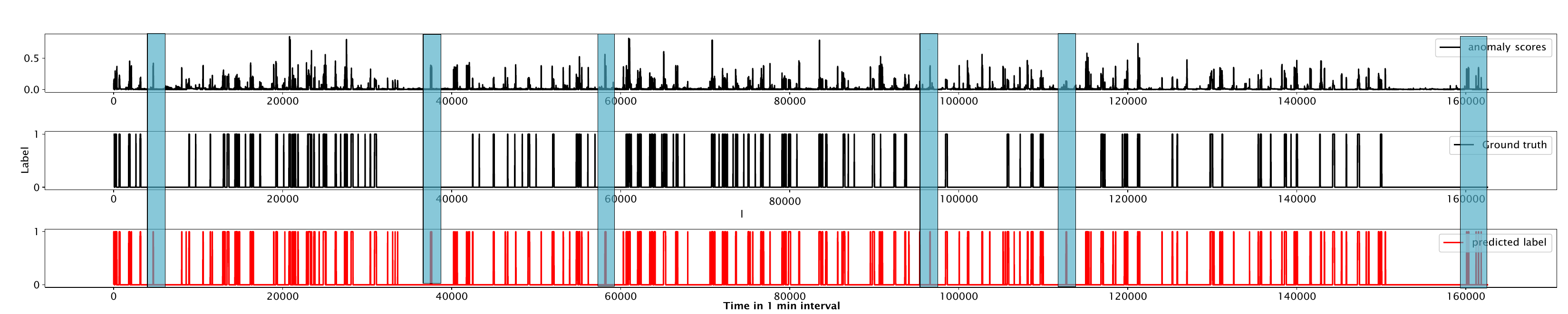}}\\
\caption{Examples of the anomaly scores, predicted labels of our M-TR model and ground Truth labels for user 1 and user 5.}
\label{fig:detections}
\end{figure*}

\begin{table}[]
\centering
\caption{Mean of F1 and ROC of the anomaly detection for different models.}
\label{Tab:mean_results}
\begin{tabular}{lll}
	\toprule
	\scriptsize
	\textbf{Model} & \multicolumn{2}{c}{\textbf{Mean}} \\
	& \multicolumn{1}{c}{{F1}} & \multicolumn{1}{c}{{ ROC}} \\
	\toprule
	LTSM-NDT & 0.429 & 0.695 \\
	DAGMM & 0.415 & 0.668 \\
	USAD & 0.121 & 0.530 \\
	MSCRED & 0.656 & 0.828 \\
	CAE\_M & 0.232 & 0.731 \\
	GDN & 0.091 & 0.514 \\
	MTAD\_GAT & 0.239 & 0.735 \\
	TranAD & 0.311 & 0.605 \\
	\textbf{Online M-TR} & \cellcolor[HTML]{8EA9DB}\textbf{0.844} & \cellcolor[HTML]{8EA9DB}\textbf{0.910}\\
	\toprule
\end{tabular}

\end{table}
\textbf{Anomaly Diagnosis}. This section discusses the anomaly detection performance of our M-TR model for some users of the testset. \Fig~\ref{fig:detections} shows the anomaly scores generated by our M-TR model for user 1 and user 3, the ground truth labels, and the predicted labels. The shaded area shows where our model consistency generates the wrong label for some interval. We argue that the wrong label is due to the model being unable to generate proper anomaly scores and also the threshold value that has been set by SPoT. 
This can be seen by ROC score, which is not sensitive to the threshold value. Despite that, our model performs very well in an online manner.

\section{Conclusion}
\label{Sec:conclusion}
This paper proposed an unsupervised online anomaly detection technique for EV charging identification from smart meter data. This paper considers the streaming of smart meter data and aims to detect the presence of EV charging from this smart meter data in an online fashion. We proposed a memory-based transformer network (M-TR) that leverages coarse-scale historical information using local and global transformer encoders and decoder networks. The M-TR generates anomaly scores based on the reconstructed local memory, and we proposed the use of SPOT to define the anomaly threshold dynamically. The proposed M-TR was benchmarked against several state-of-the-art unsupervised learning models and anomaly detection models and showed superior performance in terms of $F_1$ and ROC scores. Also, we showed that the proposed model can run in real-time with an execusion time of 1.2 sec. for 1-minute readings, and can be applied for real-time application.


%



\ifCLASSOPTIONcaptionsoff
\newpage
\fi



\bibliographystyle{IEEEtran}
\bibliography{references}

\appendices
\newpage
\end{document}